\documentclass[a4paper,11pt]{article}
\pdfoutput=1
\usepackage{jheppub}
\usepackage[T1]{fontenc} 

\newcommand{\tr}{\mathrm{tr}}
\newcommand{\rmd}{\mathrm{d}}
\newcommand{\rme}{\mathrm{e}}
\newcommand{\rmi}{\mathrm{i}}
\newcommand{\balpha}{\bar{\alpha}}
\newcommand{\bbeta}{\bar{\beta}}

\newcommand{\bpt}{{\boldsymbol{p}_\perp}}

\newcommand{\bxt}{{\boldsymbol{x}_\perp}}
\newcommand{\byt}{{\boldsymbol{y}_\perp}}
\newcommand{\bxtp}{{\boldsymbol{x}_\perp^\prime}}
\newcommand{\bytp}{{\boldsymbol{y}_\perp^\prime}}
\newcommand{\LQCD}{\Lambda_{\text{QCD}}}
\newcommand{\Nc}{N_{\mathrm{c}}}
\newcommand{\Qs}{Q_{\mathrm{s}}}

\newcommand{\tf}{{T_{\mathrm{F}}}}
\newcommand{\tfa}{{T_{\mathrm{F}}^\ast}}
\newcommand{\calT}{\mathcal{T}}
\newcommand{\GeV}{\;\text{GeV}}
\newcommand{\fm}{\;\text{fm}}
\newcommand{\pmax}{p_{\text{max}}}
\newcommand{\calO}{\mathcal{O}}

\title{\boldmath General formulae for dipole Wilson line correlators
       with the Color Glass Condensate}

\author[a]{Kenji Fukushima}
\author[b]{Yoshimasa Hidaka}
\affiliation[a]{Department of Physics, The University of Tokyo,\\
                7-3-1 Hongo, Bunkyo-ku, Tokyo 113-0033, Japan}
\affiliation[b]{Theoretical Research Division, Nishina Center,
                RIKEN,\\ 2-1 Hirosawa, Wako, Saitama 351-0198, Japan}
\emailAdd{fuku@nt.phys.s.u-tokyo.ac.jp}
\emailAdd{hidaka@riken.jp}

\abstract{
  We present general formulae to compute Wilson line correlators with
  the Color Glass Condensate described by the McLerran-Venugopalan
  model.  We explicitly construct a complete and non-orthogonal set of
  color-singlet bases and write matrix elements down, so that the
  exponential of the matrix leads to the Wilson line correlators.  We
  further develop a systematic perturbative expansion of dipole Wilson
  line correlators in terms of $1/\Nc$ where $\Nc$ is the color
  number.  As a phenomenological application we calculate the flow
  harmonics $v_n\{m\}$ in the dipole model and discuss the $\Nc$
  scaling.
}

\begin{document} 
\maketitle
\flushbottom

\section{Introduction}
\label{sec:intro}

Quantum Chromodynamics (QCD) has a special property called the
asymptotic freedom that implies that the strong coupling constant runs
to a smaller value in parton reactions involving harder momentum
scales.  In this way the perturbative calculation of QCD is a reliable
theoretical description in high-energy nuclear physics.  In reality,
however, a na\"{i}ve perturbative expansion breaks down for
diffractive type reactions in which exchanged momenta are not
necessary hard as compared to the collision energy.  Then, it is
indispensable to make a resummation over large logarithmic enhancement
factors which appear kinematically.  Such a resummation leads to a
picture of increasing parton or gluon density with increasing
scattering energy, and, one would anticipate that the parton density
would eventually enter a new regime where the effect of parton
overlapping is significant.  In such a regime of dense partons, QCD is
still perturbative in a sense that the running coupling constant $g$
is small, but is highly non-linear due to the gluon amplitude $A^\mu$
as large as $\sim 1/g$ [and thus $gA^\mu \sim \calO(1)$].  This
non-linear dynamics is a manifestation of the parton or gluon
saturation (see
Refs.~\cite{Mueller:1993rr,Mueller:1994jq,Balitsky:1995ub,Balitsky:2001re,Kovchegov:1999yj}
for pioneering extensions including the non-linearity, see also
Refs.~\cite{Gelis:2012ri,Blaizot:2016qgz} for recent reviews).

The virtue of the gluon saturation is that physical observables
exhibit universal behavior in terms of scaling variables, from which
the saturation momentum, $\Qs$, can be experimentally
fixed~\cite{Stasto:2000er}.  The theoretical framework in the
saturation regime to compute physical observables as functions of
$\Qs$ has been well established and known as the Color Glass
Condensate
(CGC)~\cite{McLerran:1993ni,McLerran:1993ka,McLerran:1994vd}.  The CGC
effective theory is elegantly formulated in the renormalization group
language~\cite{JalilianMarian:1997jx,JalilianMarian:1997gr,JalilianMarian:1997dw},
in which the soft gluons are given by the classical Yang-Mills fields
from the hard parton color source~\cite{Kovchegov:1996ty} whose
transverse density is characterized by $\Qs^2$.  The non-linear
quantum evolution equation for the color source distribution is known
as the JIMWLK equation named after the authors of the pioneering
works~\cite{JalilianMarian:1997jx,JalilianMarian:1997gr,JalilianMarian:1997dw,Iancu:2000hn,Ferreiro:2001qy}.
Although solving the JIMWLK equation demands huge computational
resources (see, for example, Ref.~\cite{Weigert:2000gi} for the
reformulation using the Langevin equation, and
Refs.~\cite{Rummukainen:2003ns,Lappi:2015fma} for numerical
simulations), a Gaussian approximated solution has been
derived~\cite{Iancu:2002aq,Lappi:2012nh}.  The simplest CGC model in
the Gaussian approximation is commonly called the McLerran-Venugopalan
(MV) model, which is often used as an initial input for the JIMWLK
evolution, or, this model itself could provide us with a good
description of the qualitative features of high-energy QCD processes.

One of the most interested and testable quantities calculated in the
CGC framework is the particle
production~\cite{Kovner:2001vi,Kovchegov:2001sc,Blaizot:2004wu,Blaizot:2004wv}
and
correlation~\cite{Baier:2005dv,Marquet:2007vb,Fukushima:2008ya,Albacete:2010pg,Kovner:2016jfp},
including electromagnetic
probes~\cite{Gelis:2002ki,Gelis:2002fw,JalilianMarian:2004er,Benic:2016yqt,Benic:2016uku}.
In the MV model various correlations among gluons and quarks have been
discussed, and these predictions are to be compared to the
experimental data of jets and hadron correlations.  One of the most
well known examples is the CGC picture to understand the ridge
structure seen in the di-jet or di-hadron correlations in rapidity
space~\cite{Dumitru:2008wn,Dumitru:2010iy}.  Thus, it would be a
natural extension to apply the CGC-based calculations to account for
the collective behavior in small size systems.  As the collision
energy grows up, small size systems such as the $p$-$A$
(proton-nucleus) and even the $p$-$p$ collisions may look similar to
the $A$-$A$ collision, yielding systematic patterns of the flow
observables (see Ref.~\cite{Loizides:2016tew} for an experimental
overview).  The final state interaction might be still responsible for
the flow observables, but at the same time, one must also estimate the
initial state effect to quantify which of the initial and the final
state interactions is more important.  Recently, within the dipole
model for the particle production the systematics of the higher order
flow observables, $v_n\{m\}$ (i.e.\ $n$-th harmonics of the
$m$-particle flow) has been quantified in the glasma graph
approximation~\cite{Skokov:2014tka,Lappi:2015vta} and in the full MV
model~\cite{Dusling:2017dqg,Dusling:2017aot}.

Generally, in the CGC model calculations, the most time-consuming part
in numerics is the numerical computation of the expectation value of
the Wilson lines.  In the presence of the CGC background fields
corresponding to the saturated soft gluons, one needs to take account
of multiple scatterings which amount to the Wilson line in the eikonal
approximation.  It has been repeatedly discussed how to compute the
Wilson line correlators efficiently in separate contexts (see
Ref.~\cite{Shi:2017gcq} for a very recent work toward general formulae
for the Wilson line correlators, for instance).  Therefore, it would
be quite useful to establish a prescription to compute the Wilson line
correlators with some generality but in a fairly straightforward way.
There are several preceding works along these lines from an early
attempt in Ref.~\cite{Fukushima:2007dy} to a very recent reformulation
in Ref.~\cite{Dumitru:2017cwt} on top of an explicit calculation in
Ref.~\cite{Shi:2017gcq}.  The purpose of the present work is to
translate the powerful method of Ref.~\cite{Fukushima:2007dy} to a
more specific physics problem in a more handy form.  In particular,
the setup in Ref.~\cite{Fukushima:2007dy} was too general and it was
not clear how to utilize the formulae for phenomenological
applications.  In the present work, hence, we focus on the special
type of correlation function in terms of the
\textit{dipole operators} in the color fundamental representation,
which is the basic building block in the dipole
model~\cite{Kowalski:2006hc}.  For $n$-point dipole correlations the
problem is reduced to the exponentiation of $n!\times n!$ matrices.
The calculation of the exponential of the matrix is numerically
feasible but it is sometimes insightful to perform the large-$\Nc$
expansion, especially to identify the $\Nc$ scaling of observables.
We emphasize that our formulae take a quite convenient form for the
large-$\Nc$ expansion and we will demonstrate the expansion explicitly
up to the order where the first nonzero flow cumulants appear.  We
check the validity of the large-$\Nc$ formulae by comparing to the
exact results known for $n=2$ with $\Nc=3$.  For $n\ge 4$ we need to
be very careful of the correct large-$\Nc$ counting and we will argue
the subtlety in the numerical analysis.

\section{Master formulae}
\label{sec:master}

The Wilson line correlators that we calculate in this paper are
expressed in analogy to quantum mechanics as
follows~\cite{Fukushima:2007dy}:
\begin{equation}
  \label{eq:Wilson}
  \biggl\langle \prod_{i=1}^n U(\bxt_i)_{\beta_i\alpha_i}
  U^\ast(\byt_i)_{\bbeta_i\balpha_i} \biggr\rangle
  = \exp\bigl[-(H_0 + V)\bigr]_{\beta_1\bbeta_1 \cdots \beta_n\bbeta_n;
    \alpha_1\balpha_1 \cdots \alpha_n\balpha_n} \,,
\end{equation}
where the \textit{free Hamiltonian}, $H_0$, is defined as
\begin{equation}
  \label{eq:hamiltonian}
  H_0 = \Qs^2 \frac{2\Nc}{\Nc^2 - 1} L(0,0) \biggl[ \sum_{i=1}^n
    \bigl( \tf_i^a - \tfa_{\bar{i}}^a \bigr) \biggr]^2\,.
\end{equation}
Here, we take the summation over $a$ in group space (which is always
implicitly assumed).  In this paper we limit our considerations to the
fundamental representation only and $\tf_i^a$'s represent the elements
of su($\Nc$) algebra in the fundamental representation, while the
formula is valid for any representation.  Here, $L(0,0)$ is
quadratically divergent, so that only color states that have zero
eigenvalues of $H_0$ can make finite contributions to the Wilson line
correlators [and thus, the detailed definition of $L(0,0)$ is not
important here; see Ref.~\cite{Fukushima:2007dy} for the explicit form
of $L(0,0)$].  We note that the definition of $\Qs$ is slightly
different from Refs.~\cite{Dusling:2017dqg,Dusling:2017aot} and we
will come to this point later when we apply our formulae for the flow
observables.  For the moment, Eq.~\eqref{eq:hamiltonian} is understood
as a definition of $\Qs$ in our convention.  The color matrix in the
\textit{interaction part} is
\begin{equation}
  \label{eq:potential}
  \begin{split}
    V = -\Qs^2 \frac{2\Nc}{\Nc^2 - 1} \biggl\{
    & \sum_{i>j}^n \Bigl[ \tf_i^a \tf_j^a \Gamma(|\bxt_i-\bxt_j|)
      + \tfa_{\bar{i}}^a \tfa_{\bar{j}}^a \Gamma(|\byt_i-\byt_j|)
      \Bigr] \\
    & - \sum_{i,j=1}^n \tf_i^a \tfa_{\bar{j}}^a
    \Gamma(|\bxt_i-\byt_j|) \biggr\} \,.
  \end{split}
\end{equation}
In the above the common building block, $\Gamma(\bxt)$, is defined in
the CGC formalism by the following integral:
\begin{equation}
  \Gamma(|\bxt|) := 2g^4 \int_{\Lambda}^\infty
  \frac{\rmd k}{2\pi}\, \frac{1}{k^3}\, \Bigl[ 1
    - J_0(k|\bxt|) \Bigr]
  \simeq -\frac{g^4}{4\pi} |\bxt|^2
  \ln \bigl(|\bxt| \bar{\Lambda}\bigr)\,,
\end{equation}
where $J_{\alpha}(x)$ is the Bessel function of the first kind, 
$\Lambda$ is an infrared (IR) cutoff of order of $\LQCD$ and we
introduced a shorthand notation as
$\bar{\Lambda}:=\tfrac{1}{2}\Lambda \rme^{\gamma-1}$.  This
approximated expression has undesirable behavior in the IR region
especially when $|\bxt|\bar{\Lambda}<1$.  We cure this IR problem,
according to Ref.~\cite{Dusling:2017aot}, by introducing another
regulator as
\begin{equation}
  \Gamma(|\bxt|) \simeq \frac{g^4}{4\pi} |\bxt|^2
  \ln \biggl( \frac{1}{|\bxt| \bar{\Lambda}} + \rme \biggr)\,.
\end{equation}
We will adopt this regularized approximation for $\Gamma(\bxt)$ in
our numerical calculations later in Sec.~\ref{sec:validity}.

Now we shall sketch our strategy to proceed to concrete calculations
of the Wilson line correlators using Eq.~\eqref{eq:Wilson}.  We will first
identify all the color-singlet bases with which $H_0$ vanishes.  For
the $n$-th power product of $U$ and $U^\ast$, there are $2n$ Wilson
lines, and then the number of independent color-singlet bases should
be $n!$, as we will explain in the next section.  We can easily
construct the color index structures by taking the permutations.
After confirming that $H_0$ surely vanishes with such bases, next, we
will consider the matrix elements of $V$.  In general it is not an
easy task to find analytical expressions for the eigenvalues of $V$.
Nevertheless, it is known that the large-$\Nc$ approximation works at
good quantitative accuracy, and we will systematically make an
expansion in power of $1/\Nc$ to find analytical expressions.

\section{Color singlet bases}
\label{sec:singlet}

Because of the dipole-type structures of the Wilson lines it is easy
to find all the combinations of the color singlet indices.  One
trivial singlet is immediately found as
\begin{equation}
  \langle\{\alpha\};\{\bar{\alpha}\}|s_0\rangle
  := \Nc^{-n/2}\,
  \delta_{\alpha_1\bar{\alpha}_1}\cdots\delta_{\alpha_n\bar{\alpha}_n}\,.
\end{equation}
Clearly all the permutations are possible, i.e.\ we introduce
$|s_p\rangle$ as a permutation of $|s_{0}\rangle$.  For this purpose let
us introduce the symmetry group $S_n$ with elements $\pi_p$ that
denotes a permutation as
\begin{equation}
  \pi_p = \begin{pmatrix}
    1 & 2 & \cdots & n \\
    p_1 & p_2 & \cdots & p_n
  \end{pmatrix} \,.
\end{equation}
Then,
\begin{equation}
  \langle\{\alpha\};\{\bar{\alpha}\}|s_p\rangle
  := \langle\{\pi_p \alpha\};\{\bar{\alpha}\}|s_0\rangle
  = \Nc^{-n/2}\,
  \delta_{\alpha_{p_1}\bar{\alpha}_1}\cdots\delta_{\alpha_{p_n}\bar{\alpha}_n}\,.
\end{equation}
We note that any $\pi_p$ can be expressed as a product of cycles. 
For example, it is easy to confirm the following
relation,
\begin{equation}
  \begin{pmatrix}
    1 & 2 & 3 & 4 & 5 & 6 \\
    5 & 1 & 3 & 6 & 2 & 4
  \end{pmatrix}
    = (152)\cdot (46) \,,
\end{equation}
where one-cycles, i.e.\ $(3)$ in the above case, are trivial and not
explicitly written.  Let us see several useful formulae for later
calculation checks.  Because of the well-known relation,
\begin{equation}
  2\tf^a_{\beta_i\alpha_i}\tf^a_{\beta_j\alpha_j}
  = \delta_{\beta_i\alpha_j}\delta_{\beta_j\alpha_i}
  - \frac{1}{\Nc}\delta_{\beta_i\alpha_i}\delta_{\beta_j\alpha_j} \,,
\end{equation}
we can easily prove that
\begin{equation}
  \langle\{\beta\};\{\bar{\beta}\}|2\tf^a_i\tf^a_j|s_p\rangle
  = \langle\{(i\,j)\pi_p \beta\};\{\bar{\beta}\}|
  s_0\rangle
  - \frac{1}{\Nc}\langle\{\pi_p\beta\};\{\bar{\beta}\}|s_0\rangle\,,
\end{equation}
which is schematically expressed as
\begin{equation}
  \label{eq:formula_tftf}
  2\tf_i^a\tf_{\bar{j}}^a \,\cdot\,\pi_p
  = \biggl[ (i\,j) - \frac{1}{\Nc} \biggr]\,\cdot \pi_p\,.
\end{equation}
For the complex conjugate, because $\tf^a$'s are Hermitean, the
following should hold:
\begin{equation}
  2\tfa^a_{\bar{\beta}_i\bar{\alpha}_i}\tfa^a_{\bar{\beta}_j\bar{\alpha}_j}
  = \delta_{\bar{\alpha}_j\bar{\beta}_i}\delta_{\bar{\alpha}_i\bar{\beta}_j}
  - \frac{1}{\Nc}\delta_{\bar{\alpha}_i\bar{\beta}_i}
  \delta_{\bar{\alpha}_j\bar{\beta}_j} \,,
\end{equation}
which implies
\begin{equation}
  \langle\{\beta\};\{\bar{\beta}\}|2\tfa^a_{\bar{i}}\tfa^a_{\bar{j}}|s_p\rangle
  = \langle\{\pi_p \beta\};\{(i\, j)\bar{\beta}\}|
  s_0\rangle
  - \frac{1}{\Nc}\langle\{\pi_p\beta\};\{\bar{\beta}\}|s_0\rangle\,.
\end{equation}
Thanks to the index structure of $|s_0\rangle$, we readily see;
$\langle\{\pi_p \beta\};\{(i\,j)\bar{\beta}\}|s_0\rangle
=\langle\{\pi_p (i\,j)^{-1}\beta\};\{\bar{\beta}\}|s_0\rangle$ and
trivially $(i\,j)^{-1}=(i\,j)$, so that we can again give a schematic
representation as
\begin{equation}
  \label{eq:formula_tfatfa}
  2\tfa^a_{\bar{i}} \tfa^a_{\bar{j}} \,\cdot \pi_p
  = \pi_p \,\cdot \biggl[ (i\,j) - \frac{1}{\Nc} \biggr] \,.
\end{equation}
Also, another useful formula is
\begin{equation}
  2\tf^a_{\beta_i\alpha_i} \tfa^a_{\bar{\beta}_j\bar{\alpha}_j}
  = \delta_{\beta_i\bar{\beta}_j}\delta_{\alpha_i\bar{\alpha}_j}
  - \frac{1}{\Nc}\delta_{\beta_i\alpha_i}
  \delta_{\bar{\beta}_j\bar{\alpha}_j}\,.
\end{equation}
In this case, if $p_j=i$, the contract of
$\delta_{\beta_i\bar{\beta}_j}\delta_{\alpha_i\bar{\alpha}_j}
=\delta_{\beta_{p_j}\bar{\beta}_j}\delta_{\alpha_{p_j}\bar{\alpha}_j}$
in the above and $\delta_{\alpha_{p_j}\bar{\alpha}_j}$ in
$|s_p\rangle$ makes $\Nc\delta_{\beta_{p_j}\bar{\beta}_j}$.  For
$p_j\neq i$ the contract of
$\delta_{\beta_i\bar{\beta}_j}\delta_{\alpha_i\bar{\alpha}_j}$ and
$\delta_{\alpha_{p_j}\bar{\alpha}_j}\delta_{\alpha_i\bar{\alpha}_{p^{-1}_i}}$
together with
$\delta_{\beta_{p_j}\alpha_{p_j}}\delta_{\bar{\beta}_{p^{-1}_i}\bar{\alpha}_{p^{-1}_i}}$
(which is always implicitly taken for indices not involved in $\tf^a$
nor $\tfa^a$) gives
$\delta_{\beta_i\bar{\beta}_j}\delta_{\beta_{p_j}\bar{\beta}_{p^{-1}_i}}
=(i\, p_j)\,\delta_{\beta_{p_j}\bar{\beta}_j}
\delta_{\beta_i\bar{\beta}_{p^{-1}_i}}$,
where $p^{-1}_i$ indicates an index that satisfies
$p_{p^{-1}_i}=i$.  Therefore, we establish the following schematic
relations:
\begin{equation}
  \label{eq:formula_tftfa}
  2\tf^a_i \tfa^a_j\, \pi_p = \begin{cases}
    \displaystyle \biggl( \Nc-\frac{1}{\Nc} \biggr)\,\cdot \pi_p
    & \text{for $p_j=i$} \\[1em]
    \displaystyle \biggl[ (i\,p_j) - \frac{1}{\Nc} \biggr]\, \cdot \pi_p
    & \text{for $p_j\neq i$}
  \end{cases}
\end{equation}
In what follows below, we will utilize the
formulae~\eqref{eq:formula_tftf}, \eqref{eq:formula_tfatfa}, and
\eqref{eq:formula_tftfa} to perform calculations with a compact
notation.

It would be an instructive check to see how $H_0|s_p\rangle=0$ is
satisfied for all $|s_p\rangle$.  For this purpose we first need to
expand the second-order Casimir operator in $H_0$ as
\begin{equation}
  \biggl[ \sum_{i=1}^n \bigl( \tf_i^a - \tfa_{\bar{i}}^a \bigr) \biggr]^2
  = \sum_{i=1}^n \bigl( \tf_i^{a\,2} + \tfa_{\bar{i}}^{a\,2} \bigr)
  + 2\sum_{i>j}^n \bigl( \tf_i^a\tf_j^a
  + \tfa_{\bar{i}}^a\tfa_{\bar{j}}^a \bigr)
  - 2\sum_{i,j=1}^n \tf_i^a\tfa_{\bar{j}}^a \,.
\end{equation}
Here, again, we note that the summation over $a$ is always implicitly
assumed.   The first term is nothing but the Casimir operator, so that
it is simply given by
\begin{equation}
  \label{eq:casimir}
  \sum_{i=1}^n \bigl( \tf_i^{a\,2}+\tfa_{\bar{i}}^{a\,2} \bigr)
  = n\biggl( \Nc - \frac{1}{\Nc} \biggr)\,,
\end{equation}
according to the su($\Nc$) algebra.  Using the
formulae~\eqref{eq:formula_tftf}, \eqref{eq:formula_tfatfa}, and
\eqref{eq:formula_tftfa}, we can simplify the rest (if applied to
$|s_p\rangle$) as
\begin{align}
  &\biggl[ 2\sum_{i>j}^n \bigl( \tf_i^a\tf_j^a
    + \tfa_{\bar{i}}^a\tfa_{\bar{j}}^a \bigr)
    -2 \sum_{i,j=1}^n \tf^a_i \tfa^a_{\bar{j}} \biggr]\,\cdot \pi_p \notag\\
  &= \sum_{i>j}^n \biggl\{ \biggl[ (i\,j) - \frac{1}{\Nc} \biggr] \cdot\pi_p
  + \pi_p \cdot \biggl[ (i\,j) - \frac{1}{\Nc} \biggr] \biggr\}
  - \sum_{i,j=1}^n \biggl[ \Nc\delta_{i p_j} + (i\,p_j)
    (1-\delta_{i p_j})-\frac{1}{\Nc} \biggr] \, \cdot \pi_p \notag\\
  &= -n\biggl(\Nc-\frac{1}{\Nc}\biggr)\,\cdot \pi_p
  + \sum_{i>j}^n \bigl[ (i\,j)\,\cdot \pi_p + \pi_p \cdot (i\,j)\bigr]
  - \sum_{i,j=1}^n (i\,p_j)(1-\delta_{i p_j})\,\cdot \pi_p \,.
\end{align}
The first term cancels out with Eq.~\eqref{eq:casimir}.
Since $i$ and $j$ run from $1$ to $n$, we can equivalently take a
summation with respect to $j$ instead of using $p_j$, leading to
\begin{equation}
  -\sum_{i,j=1}^n (i\, p_j)(1-\delta_{i p_j})\,\cdot \pi_p
  = -\sum_{i\neq j}^n (i\, j)\,\cdot \pi_p \,.
\end{equation}
Also, we see $\pi_p\cdot (i\, j)=(p_i\,p_j)\cdot \pi_p$, which allows
us to replace $p_i$ and $p_j$ with $i$ and $j$ in the summations.
Finally we arrive at
\begin{equation}
  \sum_{i>j}^n \bigl[ (i\,j)\,\cdot \pi_p + \pi_p \cdot (i\,j)\bigr]
  - \sum_{i,j=1}^n (i\,p_j)(1-\delta_{i p_j})\,\cdot \pi_p = 0\,.
\end{equation}
This completes our confirmation of $H_0|s_p\rangle=0$ for any
$|s_p\rangle$.

For the practical calculation the most important is the evaluation of
the matrix elements of $V$.  In our formalism we can infer the matrix
elements from
\begin{align}
  V\,\cdot \pi_p &= -\Qs^2\,\frac{\Nc}{\Nc^2-1}\, \Biggl( \sum_{i>j}^n
  \biggl\{
  \Bigl[ (i\,j)-\frac{1}{\Nc}\Bigr] \Gamma(|\bxt_i-\bxt_j|)
  +\Bigl[ (p_i\,p_j)-\frac{1}{\Nc}\Bigr] \Gamma(|\byt_i-\byt_j|)
  \biggr\} \notag\\
  &\quad -\sum_{i,j=1}^n \Bigl[ \Nc\delta_{i p_j} + (i\,p_j)
    (1-\delta_{i p_j})-\frac{1}{\Nc} \Bigr] \Gamma(|\bxt_i-\byt_j|)
  \Biggr)\,\cdot \pi_p \,,
\end{align}
which can be easily verified with Eqs.~\eqref{eq:formula_tftf},
\eqref{eq:formula_tfatfa}, and \eqref{eq:formula_tftfa}.  We can
further simplify the above expression by introducing a notation for a
proper combination of $\Gamma$'s, i.e.,
\begin{equation}
  F(\bxt,\bxtp;\byt,\bytp) := \Gamma(|\bxtp-\byt|)
  +\Gamma(|\bxt-\bytp|) - \Gamma(|\bxt-\bxtp|)
  -\Gamma(|\byt-\bytp|) \,.
\end{equation}
We then define the explicit components of the matrix elements as
\begin{equation}
  V|s_p\rangle = \sum_{p'} |s_{p'}\rangle\, V_{p',\, p} \,.
\end{equation}
Here we must be careful of the fact that $|s_p\rangle$'s span a
complete set of bases but they are not orthogonal.  Using these
notations and definitions we can summarize the non-zero components as
follows:
\begin{align}
  \label{eq:Vpp}
  V_{p,\,p} &= \Qs^2 \biggl[ \sum_{i=1}^n \Gamma(|\bxt_{p_i}-\byt_i|)
    - \frac{1}{\Nc^2-1}\sum_{i>j}^n
    F(\bxt_{p_i},\bxt_{p_j};\byt_i,\byt_j) \biggr] \,,\\
  \label{eq:Vpijp}
  V_{p(i\, j),\, p} &= \Qs^2\, \frac{\Nc}{\Nc^2-1}\,
  F(\bxt_{p_i},\bxt_{p_j};\byt_i,\byt_j) \,,
\end{align}
and other matrix elements vanish.
These formulae are our central results in the present paper.  For the
actual application of the formulae, we should compute the exponential
of $V$ as seen in Eq.~\eqref{eq:Wilson}.

To understand how the formulae work, let us consider the simplest
example of $n=2$.  The matrix elements of $2\times 2$ matrix $V$
read
\begin{equation}
  V = \begin{pmatrix}
    V_{0,0} & V_{0,(21)} \\ V_{(21),0} & V_{(21),(21)}
  \end{pmatrix} = \Qs^2 \frac{2\Nc}{\Nc^2-1} \begin{pmatrix}
    \displaystyle \frac{\Nc^2-1}{\Nc} \gamma - \frac{1}{\Nc}(\alpha-\beta) &
    \gamma-\beta \\
    \alpha-\beta &
    \displaystyle \frac{\Nc^2-1}{\Nc} \alpha - \frac{1}{\Nc}(\gamma-\beta)
  \end{pmatrix} \;,
\end{equation}
where,
\begin{equation}
  \begin{split}
    2\alpha := \Gamma(|\bxt_1-\byt_2|) + \Gamma(|\byt_1-\bxt_2|)\,,\\
    2\beta  := \Gamma(|\bxt_1-\bxt_2|) + \Gamma(|\byt_1-\byt_2|)\,,\\
    2\gamma := \Gamma(|\bxt_1-\byt_1|) + \Gamma(|\bxt_2-\bxt_2|)\,.
  \end{split}
\end{equation}
It is a straightforward calculation to obtain two eigenvalues as
$\lambda_\pm=\frac{1}{2}(\tr V\pm\varphi)$ with
\begin{equation}
  \varphi := \sqrt{(\tr V)^2-4\det V} = \Qs^2 \frac{2\Nc^2}{\Nc^2-1}
  \sqrt{(\alpha-\gamma)^2 + \frac{4}{\Nc^2}(\beta-\alpha)(\beta-\gamma)}\;.
\end{equation}
Now we can express the exponential of $V$ in a simple form as
\begin{equation}
  \label{eq:n2}
  \rme^{-V} = \rme^{-\frac{1}{2}\tr V} \begin{pmatrix}
    \displaystyle \cosh\tfrac{1}{2}\varphi
    - \frac{\sinh\tfrac{1}{2}\varphi}{\varphi} (V_{0,0}-V_{(21),(21)}) &
    \displaystyle -2\frac{\sinh\tfrac{1}{2}\varphi}{\varphi} V_{0,(21)} \\[1em]
    \displaystyle -2\frac{\sinh\tfrac{1}{2}\varphi}{\varphi} V_{(21),0} &
    \displaystyle \cosh\tfrac{1}{2}\varphi
    + \frac{\sinh\tfrac{1}{2}\varphi}{\varphi} (V_{0,0}-V_{(21),(21)})
  \end{pmatrix} \,.
\end{equation}
Although the calculation machinery is rather simple, a larger $n$
would cause a huge computational cost.  Hence, we will seek for an
algorithmic expansion to approximate $\rme^{-V}$ without complicated
matrix algebra.

\section{Dipole Wilson line correlators and the large-$\Nc$ expansion}
\label{sec:dipole}

For the application for the particle production problem in the
relativistic heavy-ion
collision~\cite{Dusling:2017dqg,Dusling:2017aot}, we are specifically
interested in the correlation functions of the dipole operators.  The
definition of the dipole operator is
\begin{equation}
  D(\bxt,\byt) := \frac{1}{\Nc}\tr\bigl[ U(\bxt)\,U^\dag(\byt) \bigr]
  = \frac{1}{\Nc}\,\delta_{\beta\bar{\beta}}\delta_{\alpha\bar{\alpha}}
  \, U(\bxt)_{\beta\alpha}\, U^\ast(\byt)_{\bar{\beta}\bar{\alpha}}\,.
\end{equation}
From this form it is obvious that the $n$ dipole expectation value is
given by an matrix element of $\rme^{-V}$ evaluated with
$|s_0\rangle$, i.e.,
\begin{equation}
  \biggl\langle \prod_{i=1}^n D(\bxt_i,\byt_i)\biggr\rangle
  = \langle s_0|\, \rme^{-V}\, |s_0\rangle \,.
\end{equation}
We may be able to do a direct computation, but we can develop a more
sophisticated method assuming that $\Nc$ is large enough.  In view of
Eqs.~\eqref{eq:Vpp} and \eqref{eq:Vpijp}, the off-diagonal components
are suppressed by $1/\Nc$, so that we can avoid exponentiating $V$ but
make a systematic expansion in terms of $V_{p(ij),p}$.

For notational brevity we shall denote the diagonal and the
off-diagonal parts of $V$ as $V^{(0)}$ and $V^{(1)}$, respectively.
Then, the starting point for the systematic perturbative expansion is
the \textit{interaction picture} as in quantum mechanics expressed as
\begin{equation}
  \rme^{-V} = \rme^{-V^{(0)}}\, \calT_\tau\, \exp\biggl[
    -\int_0^1 \rmd\tau\, V^{(1)}(\tau) \biggr]\,,
\end{equation}
where $\calT_\tau$ stands for the time-ordered product in terms of
$\tau$ and the time dependent $V^{(1)}(\tau)$ in the interaction
picture is defined as
\begin{equation}
  V^{(1)}(\tau) := \rme^{\tau V^{(0)}} V^{(1)}\,
  \rme^{-\tau V^{(0)}}\,.
\end{equation}
Thus, up to the second order in $V^{(1)}$ for example, the
perturbative expansion reads
\begin{equation}
  \rme^{-V} \simeq \rme^{-V^{(0)}} \biggl[
    1-\int_0^1 \rmd\tau\, V^{(1)}(\tau) +
    \int_0^1 \rmd\tau_1 \int_0^{\tau_1} \rmd\tau_2\,
    V^{(1)}(\tau_1)\,V^{(1)}(\tau_2) \biggr] \,.
\end{equation}
Because $V^{(0)}$ is a diagonal matrix, its matrix elements,
$V_{p,p}$, are the eigenvalues of $V$.  Then, let us introduce an
eigenvector $|p\rangle$ with an eigenvalue $E_p$ for $V^{(0)}$.  That
is,
\begin{equation}
  V^{(0)}|p\rangle = E_p|p\rangle = (E_p^{(0)}+E_p^{(2)})|p\rangle\,,
\end{equation}
where we decomposed the eigenvalue according to the $1/\Nc$ order as
\begin{align}
  E_p^{(0)} &:= \Qs^2 \sum_{i=1}^n\, \Gamma(\bxt_{p_i}-\byt_i)\,,\\
  E_p^{(2)} &:= -\Qs^2 \frac{1}{\Nc^2-1} \sum_{i>j}^n
  F(\bxt_{p_i},\bxt_{p_j};\byt_i,\byt_j)
  = -\frac{1}{\Nc} \sum_{i>j}^n V_{p(ij), p}^{(1)}\,.
\end{align}
It is then easy to re-express the first perturbative correction as
\begin{align}
  -\rme^{-V^{(0)}} \int_0^1 \rmd\tau\, V^{(1)}(\tau)|p\rangle
  &= -\sum_q |q\rangle\, \rme^{-E_q} \int_0^1 \rmd\tau\,
  V_{qp}^{(1)}\, \rme^{\tau(E_q-E_p)} \notag\\
  &= \sum_q |q\rangle\, \frac{\rme^{-E_q}-\rme^{-E_p}}{E_q-E_p}\,
  V_{qp}^{(1)}\,.
\end{align}
Here, $V_{qp}^{(1)}$ is defined as
$V^{(1)}|p\rangle = \sum_q |q\rangle\, V_{qp}^{(1)}$, where we note
that this $V^{(1)}$ is an original matrix, not the one in the
interaction picture.

In the same way, we can proceed to the second perturbative correction
as
\begin{align}
  &\rme^{-V^{(0)}} \int_0^1 \rmd\tau_1 \int_0^{\tau_1} \rmd\tau_2\,
  V^{(1)}(\tau_1) V^{(1)}(\tau_2)\,|p\rangle \notag\\
  &\qquad = \sum_{q,r}|q\rangle\, \rme^{-E_q} \int_0^1\rmd\tau_1
  \int_0^{\tau_1}\rmd\tau_2\, \rme^{\tau_1(E_q-E_r) + \tau_2(E_r-E_p)}
  \, V_{qr}^{(1)}\, V_{rp}^{(1)} \notag\\
  &\qquad = \sum_{q,r}|q\rangle\, \biggl(
  \frac{\rme^{-E_q}-\rme^{-E_r}}{E_q-E_r} - \frac{\rme^{-E_q}-\rme^{-E_p}}
       {E_q-E_p} \biggr)\, \frac{V_{qr}^{(1)}\, V_{rp}^{(1)}}
       {E_r-E_p} \,.
\end{align}
At this point, we can see a general algorithm to go to arbitrary high
orders.  The next order, for example, is generated automatically via
one more iteration as
\begin{align}
  \sum_{q,s,r}|q\rangle\, \biggl[ &
    \biggl(\frac{\rme^{-E_q}-\rme^{-E_s}}{E_q-E_s}
    -\frac{\rme^{-E_q}-\rme^{-E_r}}{E_q-E_r}\biggr)
   \frac{1}{E_s-E_r} \notag  \\ 
    & - \biggl(\frac{\rme^{-E_q}-\rme^{-E_s}}{E_q-E_s}
    - \frac{\rme^{-E_q}-\rme^{-E_p}}{E_q-E_p}\biggr)
        \frac{1}{E_s-E_p} 
        \biggr]\, \frac{V_{qs}^{(1)}\, V_{sr}^{(1)}\, V_{rp}^{(1)}}
      {E_r-E_p} \,.
      \label{eq:NNLO}
\end{align}

Now, we are ready to compute $\langle s_0|\rme^{-V}|s_0\rangle$ up to
the $\Nc^{-2}$ order.  Noting that $V^{(1)}$ has a matrix element
between $p$ and $p(ij)$, we can write down
\begin{align}
  \langle s_0|\rme^{-V}|s_0\rangle \simeq
  \langle s_0|s_0\rangle \, \rme^{-E_0} & + \sum_{i>j}^n
    \langle s_0|(ij)s_0\rangle\, \frac{\rme^{-E_{(ij)}}-\rme^{-E_0}}
            {E_{(ij)}-E_0}\, V_{(ij), 0}^{(1)} \notag\\
  & + \sum_{i>j}^n \langle s_0|s_0\rangle
    \biggl(\frac{\rme^{-E_0}-\rme^{-E_{(ij)}}}{E_0-E_{(ij)}}
    + e^{-E_0} \biggr)\,
    \frac{V_{0,(ij)}^{(1)}\, V_{(ij),0}^{(1)}}{E_{(ij)}-E_0} \,.
\end{align}
As we already mentioned, $|s_p\rangle$'s are not orthogonal for
different $p$'s, and a simple calculation leads to the normalization
as $\langle s_0|s_p\rangle = \Nc^{-n+n_p}$ where $n_p$ denotes the
number of cycles of $\pi_p$.
Because the second and the
third terms are already suppressed by $\Nc^{-2}$, we can replace
$E_{(ij)}$ with $E_{(ij)}^{(0)}$ in the above expansion.  Then, we
notice that $E_{(ij)}^{(0)}$ is always accompanied by $E_0^{(0)}$,
which motivates us to introduce a new notation as
\begin{equation}
  \Delta E_{(ij)}^{(0)} := E_{(ij)}^{(0)} - E_0^{(0)} = \Qs^2 \bigl[
      \Gamma(|\bxt_j-\byt_i|) + \Gamma(|\bxt_i-\byt_j|)
    - \Gamma(|\bxt_i-\byt_i|) - \Gamma(|\bxt_j-\byt_j|) \bigr]\,.
\end{equation}
Now, by expanding $E_0$ and using the above relations, we can reach
the result from the large-$\Nc$ expansion up to the second order as
\begin{equation}
  \label{eq:final}
  \biggl\langle \prod_{i=1}^n D(\bxt_i,\byt_i)\biggr\rangle
     = \rme^{-E_0^{(0)}} \Biggl\{ 1 + \sum_{i>j}^n \Biggl( 1 -
    \frac{1-\rme^{-\Delta E_{(ij)}^{(0)}}}{\Delta E_{(ij)}^{(0)}} \Biggr)\,
    \Biggl(\frac{V_{(ij),0}^{(1)}}{\Nc}
    + \frac{V_{0,\,(ij)}^{(1)}\, V_{(ij),\,0}^{(1)}}
    {\Delta E_{(ij)}^{(0)}} \Biggr)\Biggr\}\,.
\end{equation}
This is our final expression expanded up to the $\Nc^{-2}$ order in
the MV model.

\section{Comparison to the exact answer}
\label{sec:validity}

In this section let us make a comparison between numerical results
from our expansion~\eqref{eq:final} and the exact answer.  In
particular for the $n=2$ case as we discussed around
Eq.~\eqref{eq:n2}, the full analytical expression for the dipole
Wilson line correlation is known for general $\Nc$, which provides us
with a useful benchmark to quantify the validity of the large-$\Nc$
approximation in Eq.~\eqref{eq:final}.  Because our present purpose is
to check our formulae~\eqref{eq:final}, concrete values of model
parameters are not relevant.  The coupling $g$ always appears as a
combination of $g^4 \Qs^2$, so we can take $g=1$ without loss of
generality and change $\Qs$.  We measure all variables in the unit of
$\bar{\Lambda}$ here.  This means, for example, $\Qs=5$ in this
section is actually $\Qs=5\bar{\Lambda}$, etc.

\begin{figure}
  \centering
  \includegraphics[width=0.6\textwidth]{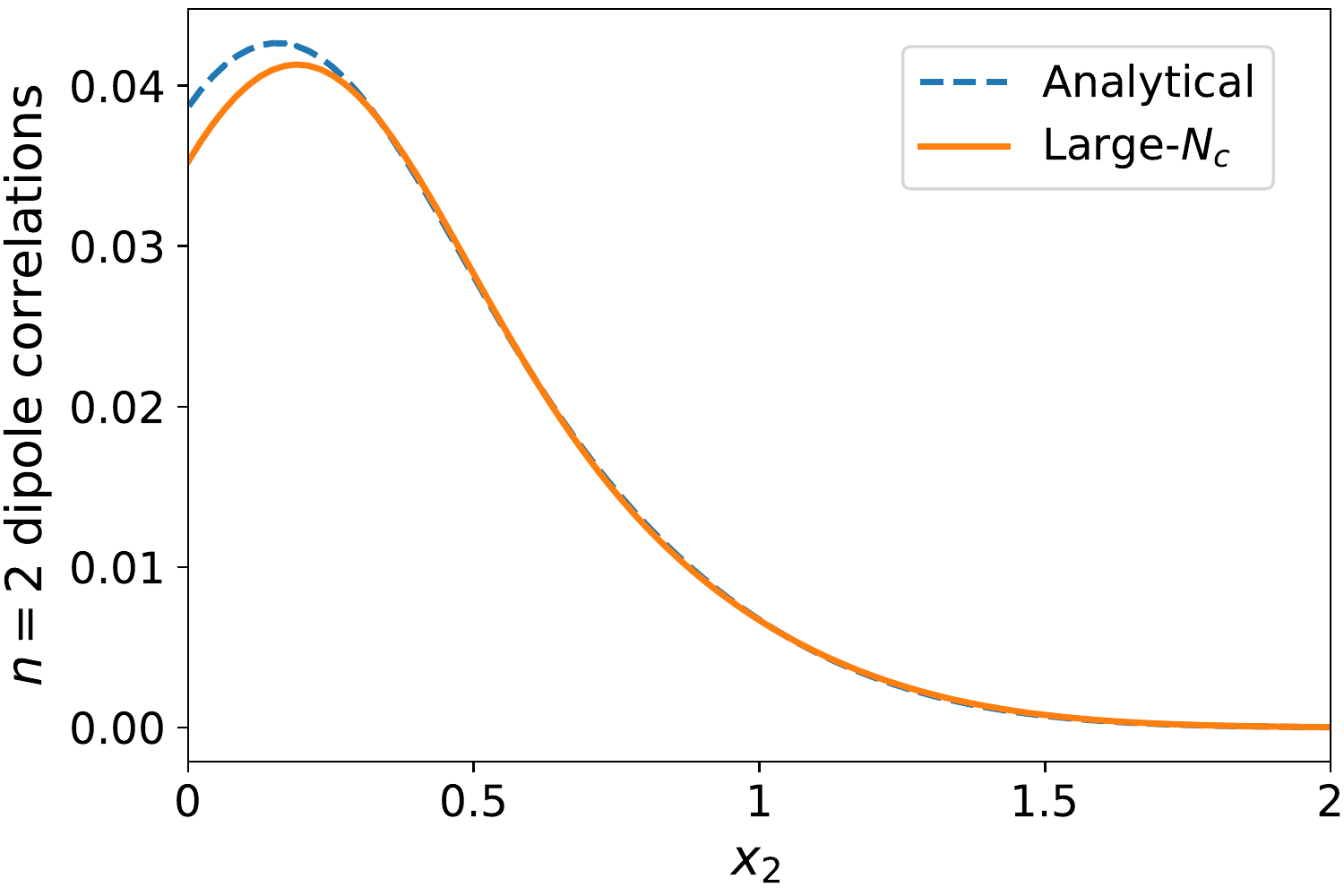}
  \caption{Validity check for the large-$\Nc$
    approximation~\eqref{eq:final} as compared to the analytically
    exact result~\eqref{eq:n2} for the $n=2$ dipole correlation with
    $\Qs=5$ at $\bxt_1=(0,0)$, $\byt_1=(0.6,0.6)$, $\bxt_2=(x_2,0)$,
    $\byt_2=(0.3,0.3)$ (in the unit of $\bar{\Lambda}$).}
  \label{fig:ntwo}
\end{figure}

Figure~\ref{fig:ntwo} shows the validity check between
Eqs.~\eqref{eq:n2} and \eqref{eq:final} for the $n=2$ dipole
correlator with $\Nc=3$.  The agreement generally depends on $\bxt_i$
and $\byt_i$, but our formulae~\eqref{eq:final} work quite well for
almost all $\bxt_i$ and $\byt_i$ as long as $\Qs$ is not too large
(when $\Qs$ is too large, the outputs are too small, and the errors
become relatively larger).  Here, in Fig.~\ref{fig:ntwo}, we chose
$\Qs=5$ and $\bxt_1=(0,0)$, $\byt_1=(0.6,0.6)$, $\bxt_2=(x_2,0)$,
$\byt_2=(0.3,0.3)$, which is intentionally chosen to make the
difference as visible as possible for the small $x_2$ region, and so,
the overall agreement is better than shown in Fig.~\ref{fig:ntwo}.

\begin{figure}
  \centering
  \includegraphics[width=0.6\textwidth]{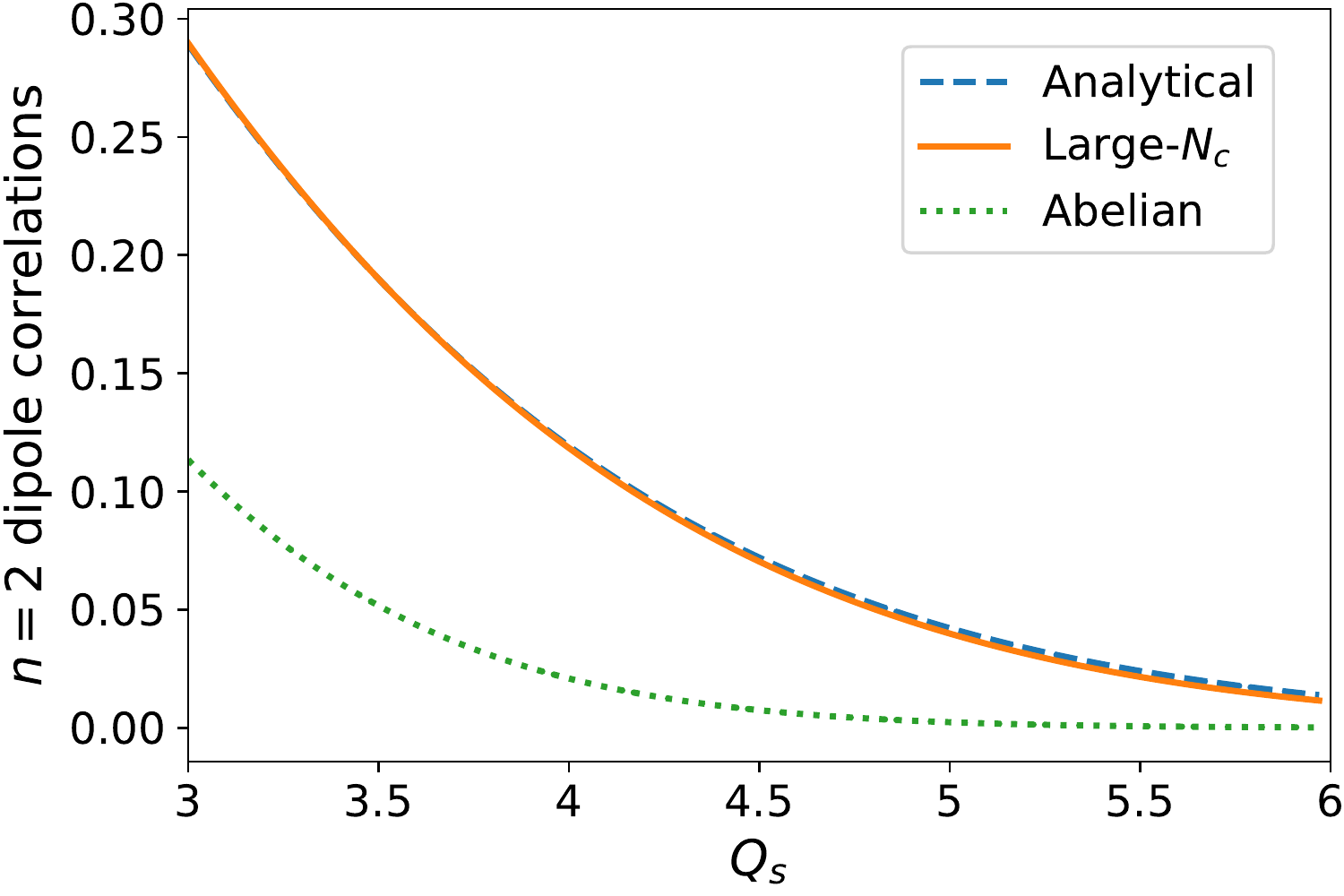}
  \caption{Validity check for the large-$\Nc$
    approximation~\eqref{eq:final} and the Abelian approximation as
    compared to the analytically exact result~\eqref{eq:n2} for the
    $n=2$ dipole correlation.  The positions are chosen as
    $\bxt_1=(0,0)$, $\byt_1=(0.6,0.6)$, $\bxt_2=(0.1,0)$,
    $\byt_2=(0.3,0.3)$.}
  \label{fig:Qs}
\end{figure}

We next see the $\Qs$ dependence of the $n=2$ dipole correlator
together with the Abelian approximation, as depicted in
Fig.~\ref{fig:Qs}.  We introduce the Abelian approximation as employed
in Ref.~\cite{Dusling:2017dqg,Dusling:2017aot} so that the $n=1$
expectation value can reproduce the exact result.  For example of
the $n=2$ case, the Abelian approximation reads
\begin{equation}
  \label{eq:abelian}
  \rme^{ -\Qs^2 [\Gamma(|\bxt_1-\byt_1|)-\Gamma(|\bxt_1-\bxt_2|)
    +\Gamma(|\bxt_1-\byt_2|)+\Gamma(|\bxt_2-\byt_1|)-\Gamma(|\byt_1-\byt_2|)
    +\Gamma(|\bxt_2-\byt_2|) ] }\,,
\end{equation}
which would agree with the exact answer in the limit of
$\bxt_1=\byt_1$ or $\bxt_2=\byt_2$ (in which the correlator reduces to
the $n=1$ one), but deviates from the exact answer for general
$\bxt_i$ and $\byt_i$.  In this sense, the expression like
Eq.~\eqref{eq:abelian} is to be regarded as an Abelianized
extrapolation from the $n=1$ answer.  Figure~\ref{fig:Qs} clearly
shows that a small disagreement magnified in Fig.~\ref{fig:ntwo} is
actually a negligibly small difference in the whole profile over
various $\Qs$.  The Abelian approximation captures a qualitative
dependence with increasing $\Qs$, while the quantitative values should
be considered as only order estimates.

\section{Flow harmonics and higher order contributions}
\label{sec:flow}

We follow the calculations of the flow observables, $v_n\{m\}$,
according to Refs.~\cite{Dusling:2017dqg,Dusling:2017aot}.  The flows
are characteristic angular distributions defined from the $m$-particle
inclusive spectra, which are in the dipole model given by
\begin{equation}
  \begin{split}
  & \frac{\rmd^m N}{\rmd^2\bpt_1 \cdots \rmd^2\bpt_m} \\
  & = \frac{1}{(4\pi^{3}B)^m} \prod_{i=1}^m \int \rmd^2\bxt_i
  \,\rmd^2\byt_i\,\rme^{-\frac{\bxt_i^2+\byt_i^2}{2B} + \rmi(\bxt_i-\byt_i)\cdot\bpt_i}
  \biggl\langle \prod_{j=1}^m D(\bxt_j, \byt_j) \biggr\rangle\,,
  \end{split}
\end{equation}
where $B$ is a dipole model parameter, which is typically of the order
of the nucleon size $\sim 1\fm\sim (0.2\GeV)^{-1}$, and we take
$\sqrt{B}=2\GeV^{-1}$.
The general analysis for the flow properties is presented in
Ref.~\cite{Borghini:2001vi} and the $n$-th moment of the $m$-particle
correlation is introduced as
\begin{equation}
  \kappa_n\{m\} := \prod_{i=1}^m \int\frac{\rmd^2\bpt_i}{(2\pi)^2}
  \, \rme^{\rmi n (-1)^{i+1}\phi_i}\,
  \frac{\rmd^m N}{\rmd^2 \bpt_1 \cdots \bpt_m}\,,
\end{equation}
where $\phi_i$ represents the azimuthal angle, i.e.\
$\bpt_i=|\bpt_i|(\cos\phi_i,\sin\phi_i)$.  Then, in the dipole model,
we can perform the momentum integrations to find the following
expression,
\begin{equation}
  \label{eq:kappa}
  \begin{split}
    & \kappa_n\{m\} \\
    & = \frac{1}{(4\pi^{3}B)^m}  \prod_{i=1}^m
    \int\rmd^2\bxt_i\,\rmd^2\byt_i\,
    \rme^{-\frac{\bxt_i^2+\byt_i^2}{2B}} K_n^{((-1)^{i+1})}(\bxt_i-\byt_i)
    \biggl\langle \prod_{j=1}^m D(\bxt_j,\byt_j)\biggr\rangle\,.
  \end{split}
\end{equation}
Here, using the regularized generalized hypergeometric function, we
defined,
\begin{align}
  K_n^{(\pm)}(\bxt) &:= \int^{\pmax} \frac{\rmd^2\bpt}{(2\pi)^2}\,
  \rme^{\pm\rmi n \phi + \rmi \bxt\cdot\bpt} \notag\\
  &= \frac{\rmi^n \pmax^2\,\rme^{\pm\rmi n\theta}}{2\pi (n+2) n!}
  \biggl(\frac{|\bxt| \pmax}{2}
  \biggr)^n \!\! {{}_1F_2} \biggl(1+\frac{n}{2};
  \Bigl\{ 1+n,\, 2+\frac{n}{2}\Bigr\};-\frac{1}{4}|\bxt|^2\pmax^2\biggr)\,,
\end{align}
where $\theta$ is the azimuthal angle of $\bxt$.  In particular, we
will frequently use the $n=0$ function for the normalization, which is
given by
\begin{equation}
  \label{eq:K0}
  K_0^{(\pm)}(\bxt) = \tilde{\delta}_{\pmax}^{(2)}(\bxt)
  := \frac{\pmax}{2\pi|\bxt|}J_1(|\bxt|\pmax)\,.
\end{equation}
It is important to notice that $\pm$ is irrelevant for $n=0$ and there
is no angular dependence any more in $K_0^{(\pm)}(\bxt)$.  Also, we
must point out that $\tilde{\delta}_{\pmax}^{(2)}(\bxt)$ should
approach $\delta^{(2)}(\bxt)$ in the $\pmax\to\infty$ limit.

Let us first consider the case with $m=2$ using our large-$\Nc$
expansion. The $\Nc^{0}$ order term in Eq.~\eqref{eq:final} does 
not contribute to $\kappa_n\{2\}$ due to the phase factor in
$K_n^{(\pm)}(\bxt_i-\byt_i)$.
As a result, we can write $\kappa_n\{2\}$ using the $\Nc^{-2}$ order
results  as
\begin{equation}
  \kappa_n\{2\}=D_n^{(+-)}\,,
\end{equation}
where
\begin{equation}
  \begin{split}
  D_n^{(+-)} &:= \frac{1}{(4\pi^{3}B)^2} 
  \int\rmd^2\bxt_1\,\rmd^2\byt_1\,\rmd^2\bxt_2\,\rmd^2\byt_2\,
  \rme^{-\frac{\bxt_1^2+\byt_1^2+\bxt_2^2+\byt_2^2}{2B}} \\
  &\quad \times \rme^{-\Qs^2 \Gamma(\bxt_1-\byt_1)}K_n^{(+)}(\bxt_1-\byt_1)
  \,\rme^{-\Qs^2 \Gamma(\bxt_2-\byt_2)}K_n^{(-)}(\bxt_2-\byt_2) \\
  &\quad \times \biggl( 1-\frac{1-\rme^{-\Delta E_{(21)}^{(0)}}}
        {\Delta E_{(21)}^{(0)}}\biggr)\biggl(\frac{V_{(21),0}^{(1)}}{\Nc}
        +\frac{V_{0,(21)}^{(1)} V_{(21),0}^{(1)}}{\Delta
          E_{(21)}^{(0)}} \biggr)\,.
  \end{split}
\end{equation}
One could think of $D_n^{(++)}$ and $D_n^{(--)}$ in a similar manner but
they are also vanishing because of the phase factors in $K_n^{(\pm)}$.  The
last part of the integrand is a function of modulus of various
combinations of $\bxt_1$, $\bxt_2$, $\byt_1$, $\byt_2$.  Here, it is
crucially important to understand that any term in the
integrand which is factorized into a function of $|\bxt_i-\byt_i|$
alone would vanish due to the phase factors in
$K_n^{(\pm)}(\bxt_i-\byt_i)$ in the factorized integrations.

\begin{figure}
  \centering
  \includegraphics[width=0.8\textwidth]{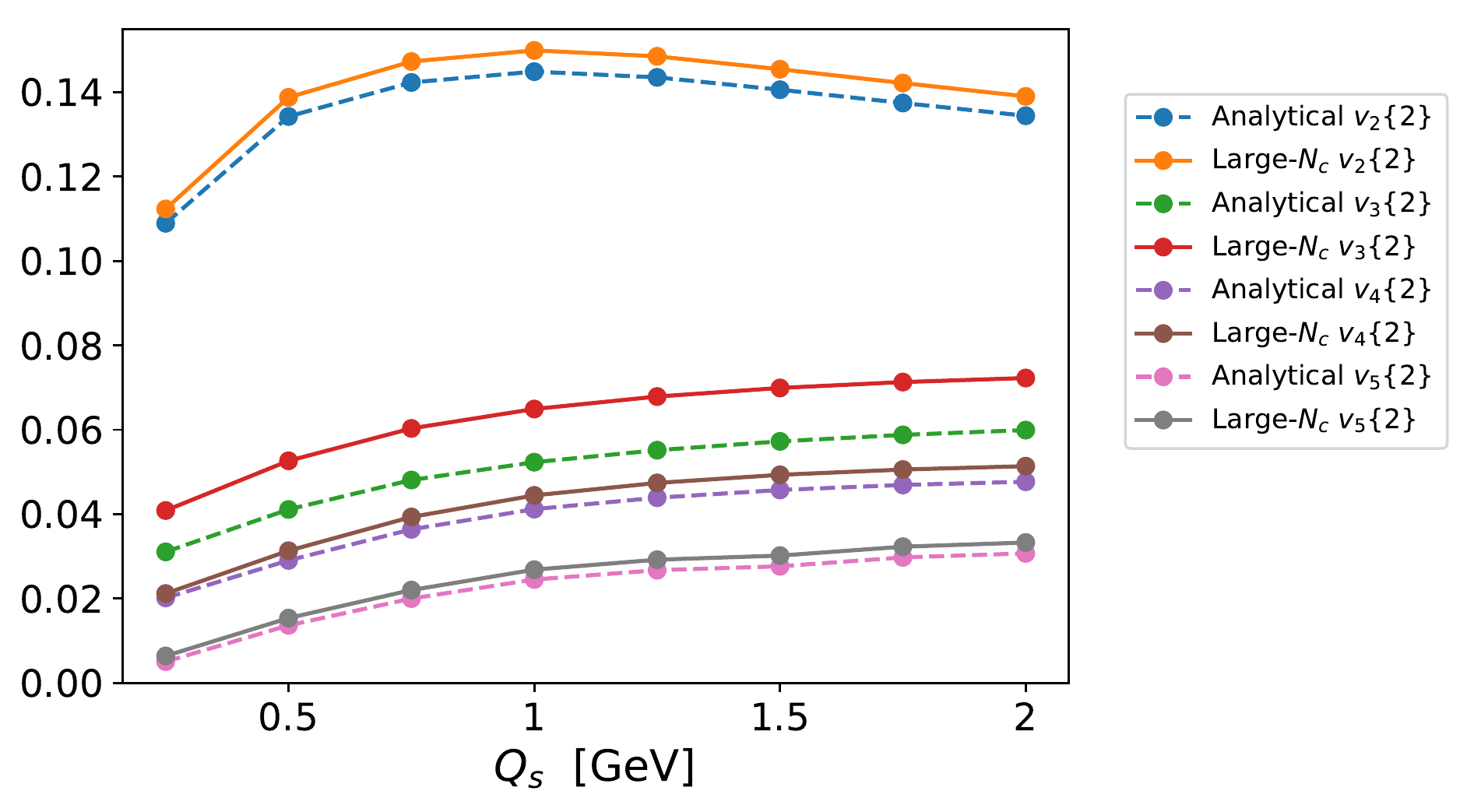}
  \caption{2-particle flow harmonics using the analytical
    (unapproximated) expression~\eqref{eq:n2} and our
    formulae~\eqref{eq:final}.  The saturation momentum $\Qs$ is the
    one defined in Eq.~\eqref{eq:newQs} as in
    Ref.~\cite{Dusling:2017aot}.}
    \label{fig:flow}
\end{figure}

Because there is no finite contribution of disconnected parts in the two particle correlation,
we can immediately compute the two particle flow harmonics,
$v_n\{2\}$, from
\begin{equation}
  v_n\{2\} = \sqrt{ \frac{\kappa_n\{2\}}{\kappa_0\{2\}} }\,,
\end{equation}
where the denominator is obtained with Eq.~\eqref{eq:K0}, i.e.\ if we
keep using the expanded expression up to the $\Nc^{-2}$ order for
later convenience, we have
\begin{equation}
  \label{eq:k02}
  \kappa_0\{2\} = D_0^2 + D_0^{(+-)}\,.
\end{equation}
Here, we defined
\begin{equation}
  D_0 :=\frac{1}{4\pi^{3}B} \int\rmd^2\bxt\,\rmd^2\byt\,
  \rme^{-\frac{\bxt^2+\byt^2}{2B}}\,
  \rme^{-\Qs^2\Gamma(\bxt-\byt)}\,\tilde{\delta}_{\pmax}^{(2)}(\bxt-\byt)\,.
\end{equation}
Of course, up to this order, keeping $D_0^{(+-)}$ in the denominator
is in principle irrelevant since it gives a higher order correction
which we should neglect.

We summarize our numerical results in Fig.~\ref{fig:flow}.  We have
performed the 8 dimensional numerical integration with respect to
$\{\bxt_i,\byt_i\}$ using the Monte-Carlo method by taking
$10^8$ sampling points.  To draw Fig.~\ref{fig:flow} we 
chose $\bar{\Lambda}=0.241\GeV$ and $\pmax=2\GeV$ in accord with
Ref.~\cite{Dusling:2017aot}.  We also note that, only in this section,
we change the definition of the saturation momentum from our original
$\Qs$ in Eqs.~\eqref{eq:hamiltonian} and \eqref{eq:potential} to new
$\bar{Q}_{\text{s}}$ defined by
\begin{equation}
  \label{eq:newQs}
  \bar{Q}_{\text{s}}^2 = \frac{1}{2\Gamma(|\sqrt{2}/\Qs|)}
\end{equation}
according to Refs.~\cite{Dusling:2017dqg,Dusling:2017aot}.
Since there is no confusion, in this section, we will omit bar and
simply denote $\Qs$ to mean $\bar{Q}_{\text{s}}$.  Then, we can make a
direct comparison of our outputs to Fig.~1 of
Ref.~\cite{Dusling:2017dqg}.  The dashed curved in Fig.~\ref{fig:flow}
must precisely reproduce Fig.~1 of Ref.~\cite{Dusling:2017dqg}.  At a
glance of our numerical calculations we see quantitatively good
agreement.  The most interesting question is how useful our
large-$\Nc$ formulae~\eqref{eq:final} can be for the $\Nc=3$ case.
The comparison between the dashed curves (full analytical results) and
the solid curves (large-$\Nc$ approximations) in Fig.~\ref{fig:flow}
concludes that the errors are of only a few (at most $\sim 5$) \%
level except for the $n=3$ case.

Next, it is intriguing to see what happens for $m=4$.  In this case,
$\kappa_n\{4\}$ involves $K_n^{(+)}(\bxt_1-\byt_1)$,
$K_n^{(-)}(\bxt_2-\byt_2)$, $K_n^{(+)}(\bxt_3-\byt_3)$, and
$K_n^{(-)}(\bxt_4-\byt_4)$.  It is then easy to understand that our
formula of the $\Nc^{-2}$ order in Eq.~\eqref{eq:final} is
insufficient to get nonvanishing contributions.  Terms of the
formula~\eqref{eq:final} are functions of, say, $\bxt_1$, $\bxt_2$,
$\byt_1$, $\byt_2$ for $i=2$ and $j=1$, and then the angle
integrations of $\bxt_3-\byt_3$ and $\bxt_4-\byt_4$ become zero.
Therefore, one permutation is not enough, but two permutations are
necessary to shuffle all $\bxt_1$, $\bxt_2$, $\bxt_3$, and $\bxt_4$;
we must go to the next $\Nc^{-4}$ order to compute a first nonzero
term in $\kappa_n\{4\}$.

It is a straightforward but tedious calculation to pick all the
$\Nc^{-4}$ order terms up from the expansion in Eq.~\eqref{eq:NNLO}.
We can slightly simplify the problem by discarding terms which do not
contribute to $\kappa_n\{4\}$.  The $\Nc^{-4}$ order terms generally
contain the product of the interaction matrix elements like
\begin{equation}
  \sim V_{0,(pq)}^{(1)} V_{(pq),(ij)(kl)}^{(1)}
  V_{(ij)(kl),(ij)}^{(1)} V_{(ij),0}^{(1)}\,,
\end{equation}
but we already saw that, if an unexchanged pair exists, the angle
integration is vanishing.  For example, if the above matrix elements
are $\sim V_{0,(21)}^{(1)} V_{(21),0}^{(1)} V_{0,(21)}^{(1)}
V_{(21),0}^{(1)}$, which itself is nonzero, the angle integrations of
$\bxt_3-\byt_3$ and $\bxt_4-\byt_4$ are zero.  Thus, among all
possible combinations of the matrix elements, there are finite
contributions only from
\begin{equation}
  \label{eq:ijkl}
  \begin{split}
  & V_{0,(43)}^{(1)} V_{(43),(0)}^{(1)}
  V_{(0),(21)}^{(1)} V_{(21),0}^{(1)}\,,\qquad
  V_{0,(42)}^{(1)} V_{(42),(0)}^{(1)}
  V_{(0),(31)}^{(1)} V_{(31),0}^{(1)}\,,\\
  & V_{0,(32)}^{(1)} V_{(32),(0)}^{(1)}
  V_{(0),(41)}^{(1)} V_{(41),0}^{(1)}\,.
  \end{split}
\end{equation}
Here, we used relations such as
$V_{(ij)(kl),(ij)}^{(1)}=V_{(kl),0}^{(1)}$,
$V_{(kl),(ij)(kl)}^{(1)}=V_{0,(ij)}^{(1)}$, etc for $(ij)\neq(kl)$.
After long calculations we arrive at a final form which turned out to
be factorized as
\begin{align}
  \kappa_n\{4\} &= \frac{1}{(4\pi^{3}B)^4} 
  \prod_{i=1}^4 \int\rmd^2\bxt_i\,\rmd^2\byt_i\,
  \rme^{-\frac{\bxt_i^2+\byt_i^2}{2B}}\,
  \rme^{-\Qs^2 \Gamma(\bxt_i-\byt_i)}K_n^{((-1)^{i+1})}(\bxt_i-\byt_i) \notag\\
  &\quad \times \frac{1}{2!}\sum_{(ij)\neq(kl)}
  \biggl( 1-\frac{1-\rme^{-\Delta E_{(ij)}^{(0)}}}
        {\Delta E_{(ij)}^{(0)}}\biggr)\biggl(\frac{V_{(ij),0}^{(1)}}{\Nc}
        +\frac{V_{0,(ij)}^{(1)} V_{(ij),0}^{(1)}}{\Delta
          E_{(ij)}^{(0)}} \biggr) \notag\\
  &\qquad\qquad\quad\; \times \biggl( 1-\frac{1-\rme^{-\Delta E_{(kl)}^{(0)}}}
        {\Delta E_{(kl)}^{(0)}}\biggr)\biggl(\frac{V_{(kl),0}^{(1)}}{\Nc}
        +\frac{V_{0,(kl)}^{(1)} V_{(kl),0}^{(1)}}{\Delta
          E_{(kl)}^{(0)}} \biggr)\,,
\end{align}
which is the full expression of the $\Nc^{-4}$ order without any
truncation like the glasma graph approximation.  This result looks
quite reasonable, but we emphasize that the complete cancellation of
$(ij)(kl)$ intermediate states with an energy denominator,
$E_{(ij)}+E_{(kl)}$, is far from trivial.
The sum with respect to $(ij)$ and $(kl)$ should run over
all the permutations of the combinations as listed in
Eq.~\eqref{eq:ijkl}.  Among all the combinations of indices,
$[(ij)=(31), (kl)=(42)]$ and $[(ij)=(42), (kl)=(31)]$ are irrelevant
because $D_n^{(++)}=D_n^{(--)}=0$ (for $n>0$) as we already pointed
out.  Therefore, the remaining four combinations of
$[(ij)=(21), (kl)=(43)]$, $[(ij)=(43), (kl)=(21)]$,
$[(ij)=(41), (kl)=(32)]$, and $[(ij)=(32), (kl)=(41)]$ lead to
\begin{equation}
  \kappa_n\{4\} = 2 \bigl[ D_n^{(+-)} \bigr]^2\,.
\end{equation}
Then, with extra terms corresponding to $D_0^{(++)}$ and $D_0^{(--)}$
which are nonzero, the normalization is written as
\begin{equation}
  \kappa_0\{4\} = D_0^4 + 6D_0^2 D_0^{(+-)}\,,
\end{equation}
up to the $\Nc^{-2}$ order in the same way as in Eq.~\eqref{eq:k02}.
Now, the cumulant is then given by
\begin{equation}
  c_n\{4\} = \frac{\kappa_n\{4\}}{\kappa_0\{4\}}
  - 2 \biggl(\frac{\kappa_n\{2\}}{\kappa_0\{2\}}\biggr)^2\,.
\end{equation}
The above quantity itself is zero in the strict order counting for
$c_n\{4\}$ up to $\Nc^{-4}$.  Thus, our conclusion is, even in the
full MV model beyond the glasma graph approximation, no connected
cumulant remains for $c_n\{4\}$ at the $\Nc^{-4}$ order.

In this way we can understand that the first connected contribution to
cumulants appears from the $\Nc^{-2m+2}$ order;  for $m=4$, thus, we
need to go to the $\Nc^{-6}$ order and then a completely nested
combination of four indices like $(43), (32), (21)$ is possible.
Therefore, the flow harmonics from the fully nested permutations must
scale as
\begin{equation}
  \label{eq:scaling1}
  v_n\{m\} \sim \Nc^{-2+2/m}\,.
\end{equation}
This is our conclusion on the $\Nc$ scaling in the full MV model.

\begin{figure}
  \centering
  \includegraphics[width=0.6\textwidth]{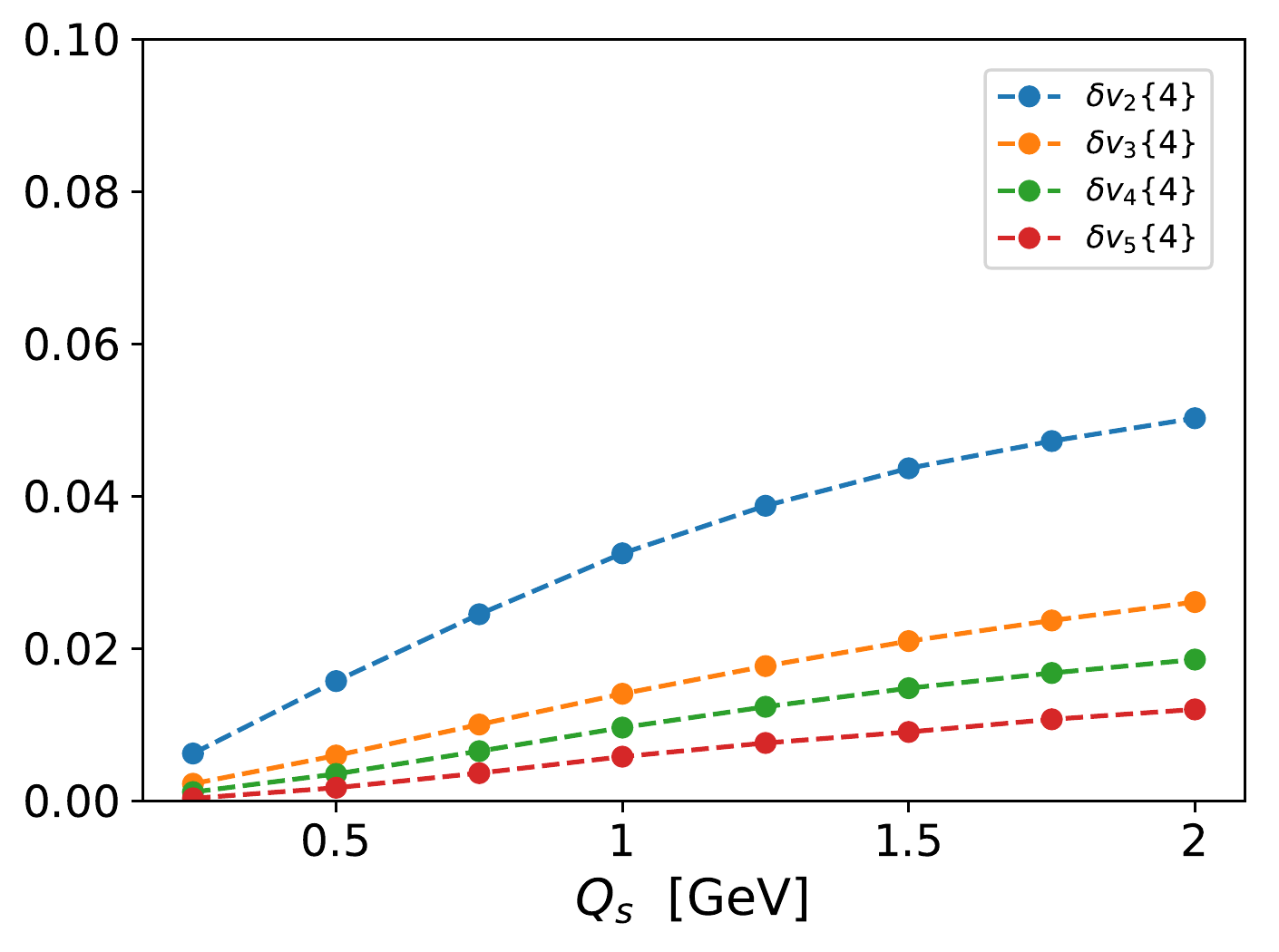}
  \caption{$\delta v_n\{4\}$ estimated from Eq.~\eqref{eq:delta} using
    $v_n\{2\}$ obtained in the large-$\Nc$ limit.  Our choice of the
    parameter is $\pmax=2\GeV$.}
    \label{fig:flowdelta}
\end{figure}

Although our conclusion of $c_n\{4\}=0$ at the $\Nc^{-4}$ order is
solid, there may be a subtle point in the large-$\Nc$ scaling due to
finiteness of $\pmax$.  If
one just uses our formulae up to the $\Nc^{-4}$ order to estimate
$\kappa_n\{2\}$, $\kappa_0\{2\}$, $\kappa_n\{4\}$, and
$\kappa_0\{4\}$, one would na\"{i}vely find that the cancellation is
incomplete at the $\Nc^{-6}$ order and a finite remainder is given by
\begin{equation}
  \delta c_n\{4\} = \frac{2[D_n^{(+-)}]^2}{D_0^4+6D_0^2D_0^{(+-)}}
  -2\biggl(\frac{D_n^{(+-)}}{D_0^2+D_0^{(+-)}}\biggr)^2
  \simeq -\frac{8[D_n^{(+-)}]^2 D_0^{(+-)}}{D_0^6}
  \simeq -\frac{8D_{0}^{(+-)}}{D_{0}^{2}}
  (v_n\{2\})^4\,.
\end{equation}
This correction is beyond the $\Nc^{-4}$ order and should be
identified as a part of the $\Nc^{-6}$ terms.  One can easily check
that this $\Nc^{-6}$ order correction is vanishing for
$\pmax\to\infty$, in which $D_0^{(+-)}\to 0$.  However, for a finite
$\pmax$ especially comparable to $\Qs$, the correction could be
sizable and the flow harmonics is modified even from the disconnected
piece.  To see this effect quantitatively, let us compute the flow
harmonics corresponding to $\delta c\{4\}$, which leads to
\begin{equation}
  \label{eq:delta}
  \delta v_n\{4\} = (-\delta c_n\{4\})^{1/4} =
  2^{3/4}  \biggl(\frac{D_{0}^{(+-)}}{D_{0}^{2}} \biggr)^{1/4}
  v_n\{2\} \,.
\end{equation}
We make a plot of $\delta v_n\{4\}$ as a function of $\Qs$ in
Fig.~\ref{fig:flowdelta}.  One must be very careful of the physical
interpretation of this correction by $\delta v_n\{4\}$.  Even though
this is non-negligible as seen in Fig.~\ref{fig:flowdelta}, the
physical origin lies in not the connected correlator but in the
normalization.  Moreover,  this normalization effect makes the $\Nc$
counting skewed to become
\begin{equation}
  \label{eq:scaling2}
  \delta v_n\{m\} \sim \Nc^{-1-2/m} \qquad (m\ge 4)\,,
\end{equation}
which starts differing from anticipated Eq.~\eqref{eq:scaling1} for
$m>4$.  Thus, to distinguish the connected contribution from the
normalization effect, one can test the $\Nc$ scaling properties as in
Eqs.~\eqref{eq:scaling1} and \eqref{eq:scaling2} and also check the
$\pmax$ dependence since $\delta v_n\{m\}$ from the normalization
effect is very sensitive to $\pmax/\Qs$ as perceived from
Fig.~\ref{fig:flowdelta}.

\section{Conclusions}
\label{sec:conclusions}

We have established general formulae and machineries to compute
$2n$-point Wilson line (or $n$ dipole) correlators with the color
group representation $(\Nc\otimes \bar{\Nc})^{n}$ in the
McLerran-Venugopalan model.  The color structure accommodates a huge
representation but the nonzero contribution to the Wilson line
correlators reduces to the $n!\times n!$ matrix, whose bases
correspond to the color singlets.  In particular, we have derived the
explicit expression of the matrix elements [see Eqs.~\eqref{eq:Vpp}
  and \eqref{eq:Vpijp}] in the color singlet bases constructed by
permutations.  The formulae are quite powerful not only for the
direct numerical evaluation of the matrix but also for the analytical
large-$\Nc$ expansion.  We have developed the systematic large-$\Nc$
expansion in a way analogous to time-dependent perturbation theory in
quantum mechanics.  We have then shown the explicit expression up to
$\Nc^{-2}$ order for the dipole correlators as given in
Eq.~\eqref{eq:final}.  As a check of the validity, we have compared
results from the exact answers and those in the large-$\Nc$ expansion
for the two-particle flow harmonics,
$v_{n}\{2\}$ $(n=2,3,4,5)$, which shows quantitatively good
agreement.  We have continued our large-$\Nc$ expansion to higher
orders to discuss the flow harmonics with more particles.  Then, we
have discovered the $\Nc$ scaling as $v_n\{m\} \sim \Nc^{-2+2/m}$ even
beyond the glasma graph approximation but in the full MV model.  We
have also pointed out that a slightly different $\Nc$ scaling could
emerge from the normalization effect at finite cutoff of the
transverse momenta of integrated particles.

Although we focused on only the dipole correlators in the present
paper, our general formulae also provide us with useful approaches to
evaluate Wilson line correlators in channels relevant for the particle
production rate in the $p$-$A$ collision generally.  In fact, not only
fundamental but also adjoint Wilson line correlators appearing in the
multi-gluon production can be derived from our results in
Eq.~\eqref{eq:Wilson} through the relation
$[U_{\text{adj}}]_{ba}(\bxt) = 2\tr [U(\bxt) \tf^{a} U^{\dag}(\bxt)
\tf^{b}] = \tf_{\alpha\bar{\alpha}}^{a}\tf_{\bar{\beta}\beta}^{b}
U_{\beta\alpha}(\bxt)U_{\bar{\beta}\bar{\alpha}}^{*}(\bxt)$.

As we emphasized, our scheme of the large-$\Nc$ expansion takes a nice
form which is easily implemented in numerical algorithms to go to
arbitrarily higher orders.  Such higher order numerical evaluations
are left as an intriguing future problem.  It would be also an
important question to think about generalizations beyond the MV
model.  Further systematic considerations on the Wilson line
correlators should deserve more investigations in the future.

\acknowledgments

The authors thank
Kevin~Dusling,
Mark~Mace,
S\"{o}ren~Schlichting,
Vladimir~Skokov,
and Raju~Venugopalan for discussions.
This work was supported by Japan Society for the Promotion of Science
(JSPS) KAKENHI Grant No.\ 15H03652, 15K13479 and 16K17716.

\bibliographystyle{JHEP.bst}
\bibliography{MV}
\end{document}